# Disclosing the nature of asymmetric interface magnetism in Co/Pt multilayers

Adriano Verna*[1,2], Paola Alippi[3], Francesco Offi[1], Gianni Barucca[4], Gaspare Varvaro[3], Elisabetta Agostinelli[3], Manfred Albrecht[5], Bogdan Rutkowski[6], Alessandro Ruocco[1], Daniele Paoloni[1], Manuel Valvidares[7] and Sara Laureti*[3]

[1] Dipartimento di Scienze, Università degli Studi Roma Tre, Via della Vasca Navale 84, I-00146 Roma, Italy

[2] ENEA-FSN-Fiss-SNI, Casaccia R.C., Via Anguillarese 301, 00123 Roma, Italy

[3] Istituto di Struttura della Materia, CNR, nM$^2$-Lab, Monterotondo Scalo (Roma), 00015, Italy

[4] Università Politecnica Delle Marche, Dipartimento SIMAU, Via Brecce Bianche, Ancona, Italy

[5] Institute of Physics, University of Augsburg, Universitätsstraße 1 Nord, D-86159 Augsburg, Germany

[6] Faculty of Metals Engineering and Industrial Computer Science, AGH University of Science and Technology, Al. A. Mickiewicza 30, 30-059 Kraków, Poland

[7] ALBA Synchrotron Light Source, E-08290 Cerdanyola del Vallès, Barcelona, Spain

*Corresponding authors – adriano.verna@enea.it; sara.laureti@ism.cnr.it

## Abstract

Nowadays a wide number of applications based on magnetic materials relies on the properties arising at the interface between different layers in complex heterostructures engineered at the nanoscale. In ferromagnetic/heavy metal multilayers, such as the [Co/Pt]$_N$ and [Co/Pd]$_N$ systems, the *magnetic proximity effect* was demonstrated to be asymmetric, thus inducing a magnetic moment on the Pt (Pd) layer that is typically higher at the top Co/Pt(Pd) interface. In this work, advanced spectroscopic and imaging techniques were combined with theoretical approaches to clarify the origin of this asymmetry both in Co/Pt trilayers and, for the first time, in multilayer systems that are more relevant for practical applications. The different magnetic moment induced at the Co/Pt interfaces was correlated to the microstructural features, which are in turn affected by the growth processes that induce a different intermixing during the film deposition, thus influencing the interface magnetic profile.

# 1. Introduction

Ferromagnetic (FM)/heavy metal (HM) multilayers (MLs), such as [X/Pt]$_N$ and [X/Pd]$_N$ (X=Co, Fe, N: bilayer repetition) thin film stacks have been extensively studied for many years owing to their strong perpendicular magnetic anisotropy (PMA) arising at the interface between the layers ($10^5 - 10^6$ J/m$^3$) [1–5], which makes them of great interest for different technological applications [6–8]. [Co/Pt(Pd)]$_N$ multilayers were firstly proposed as perpendicular recording media in high density storage technology [9] as a valid alternative to CoPt or FePt chemically ordered alloys [10–16]. To this purpose, a wide number of studies has been carried out aimed at disclosing the relationship between the PMA and the ML structural features, i.e. thickness of the sub-layers, repetition number N or crystallographic orientation [2,17]. The large tunability of their physical properties has been then exploited in spintronic devices, such as giant magnetoresistive (GMR) spin valves [8] and magnetic tunnelling junctions (MTJs) [18], where [Co/Pt(Pd)]$_N$ multilayers have been used as both the free and the reference electrodes. For this purpose, different architectures have been proposed, with the ML used as single FM electrode or in combination with suitable non-magnetic spacers (e.g., Ru, Ir, Cu, etc.) to build up synthetic antiferromagnets (SAFs) consisting of two or more FM thin films coupled antiferromagnetically [19]. More recently, the significant robustness of [Co/Pt(Pd)]$_N$ MLs along with the capability to grow on different rigid and flexible substrates while preserving their magnetic properties, have further expanded the field of applications. They have been proposed, for example, as suitable material for the preparation of freestanding micro-/nanodisks for biomedical applications [20] or as key components of flexible PMA-GMR spin valves [8] for a new generation of flexible magnetoreceptive devices with a great potential for application in human-machine interfaces [7].

Furthermore, it has been recently observed that the FM/HM interface allows stabilizing peculiar chiral spin configurations such as skyrmions [21–23] and homochiral Néel-type domain walls (DWs) [24]. The intriguing properties of this kind of noncollinear spin textures, which make DW- and skyrmion-based devices attractive for the development of advanced logic and storage technologies[25,26], are due to the interfacial Dzyaloshinskii–Moriya interaction (DMI) [27,28] which, according to the Fert/Levy model [29], results from spin-orbit coupling combined with the lack of inversion symmetry at the FM/HM interface. The electronic configuration of the materials constituting the FM/HM system determines not only the magnetization textures, but allows the electrical generation and detection of pure spin currents via the spin Hall [30] and inverse spin Hall [31] effects. Such a plethora of phenomena, which makes Pt- or Pd-based MLs key materials in heterostructures for modern spintronics applications, are widely correlated to the so-called *magnetic proximity effect* (MPE) [32], whose origin is recognized to be the hybridization between the HM-*5d* and the FM-*3d* bands, which induces a magnetic moment in heavy metals close to the Stoner instability [33].

Although many experimental [28] and theoretical [34,35] approaches have been proposed to quantitatively determine the relationship between the MPE and its impact on the DMI [24,36–38], the factors determining the strength of the proximity-induced magnetic moment in HM layers are still under investigation. To this purpose, advanced characterization techniques have been recently exploited to quantify the HM induced moment in different FM/HM architectures. In all these studies, one of the most relevant finding is an asymmetric MPE at symmetric HM/FM/HM interfaces, with a HM induced moment that is typically higher at the top interface [28,39,40]. Among the different available techniques, X-ray resonant magnetic reflectivity (XRMR) resulted effective in determining the structural and magnetic depth profile in numerous heterostructures and has been recently exploited to investigate in detail the difference between the top and the bottom Pd layers in a Pd/Co/Pd system [39]. However, the negligible difference between the structural interfacial roughness at the two FM/HM interfaces and the thickness of the two HM layers as well as the presence of a Ta buffer layer, did not allow to clearly elucidate the origin of the observed asymmetric MPE. *Mukhopadhyay* and *co-authors* have recently studied Pt(2nm)/Co(2nm)/Pt(2nm) trilayers and attributed the observed asymmetry to a different Pt crystallographic orientation induced by the presence of a Ta buffer layer [41]. However, while this might be the case for rather thick FM layers, this hypothesis is less plausible when the FM buffer film only consists of few atomic planes where the crystallographic orientation is expected to be preserved along the whole stack [42]. Although the presence of such asymmetry has been confirmed by several studies, a clear correlation between the microstructural properties and the induced magnetic moment on the HM is still lacking. Moreover, until now the investigation has been restricted to basic HM/FM/HM sandwich-like trilayers through the comparison between the bottom and the top interface. On the contrary, to our best knowledge, the occurrence of such an intriguing asymmetric effect on multilayer thin film stacks (i.e. for $(FM/HM)_N$ MLs with $N > 2$ ), which are of greater interest for practical application, has never been investigated with depth resolution.

In the present work, advanced characterization techniques and theoretical approaches were combined to thoroughly investigate the interfaces' structure at the microscopic level thus allowing a comprehensive description of the asymmetric PME induced at the Co/Pt interfaces in $[Co/Pt]_4$ MLs. Indeed, XRMR exploits the large variation of the charge and magnetic atomic scattering factors at the p→d absorption edges of transition metals; by this way, XRMR combines the spectroscopic information, X-ray absorption spectroscopy (XAS) and of X-ray magnetic circular dichroism (XMCD) with the spatial resolution due to the interference effects, allowing to reconstruct the depth profiles of elemental concentration and of magnetization with a sub-nanometer accuracy [43–45]. To support the interpretation of X-ray results, scanning transmission electron microscopy (STEM) was used for a deep structural and analytical investigation of samples at atomic spatial resolution. The correlation between the induced Pt magnetic moment and the characteristics of the interface resulting by the interdiffusion processes occurring during the ML deposition allows disclosing the origin of the asymmetric MPE, thus providing a significant contribution for the design of ML-based spintronic devices.

## 2. Methods

All samples were deposited at room temperature by DC magnetron sputtering (BESTEC UHV system, Ar sputter gas pressure of 3.5 μbar) on a thermally oxidized Si(100) wafer utilizing a Ta buffer layer for improving the adhesion of the layers deposited on top while inducing a strong (111) crystallographic texture [42]; the latter induces PMA in the in the Co/Pt multilayer system, provided that the Co thickness is fixed and limited to few monolayers [42]. The sample's stack sequence is Si(100)/SiO$_x$/Ta(30 Å)/Pt (30 Å)/Co (6 Å)/Pt (30 Å) for the reference trilayer and Si(100)/SiO$_x$/Ta(30 Å)/[Pt (22 Å)/ [Pt (8 Å)/Co (6 Å)]$_4$ /Pt (30 Å) for the multilayer, as illustrated in **Figure 1**. It is noteworthy that the choice we made in our study is to keep the thickness of Co fixed to 6 Å both for the trilayer and for the multilayer. In particular, the trilayer is made of a single Co layer that is "sandwiched" between two Pt layers (30 Å thick). Interestingly, an inverted Co/Pt/Co system has been presented in a recent paper [46]: in that case, the two Co layers with different thicknesses (30 Å and 10 Å) are characterized by orthogonal easy axes that are coupled via interlayer exchange coupling (IEC) across Pt spacer. The choice of a different thickness brings, as a main effect, to a difference in the magnetization orientation of the Co layers and in the resulting modulation of the magnetic anisotropy of the stack. Hence, this parameter represents an additional variable on the magnetic proximity effect that was carefully avoided in our study.

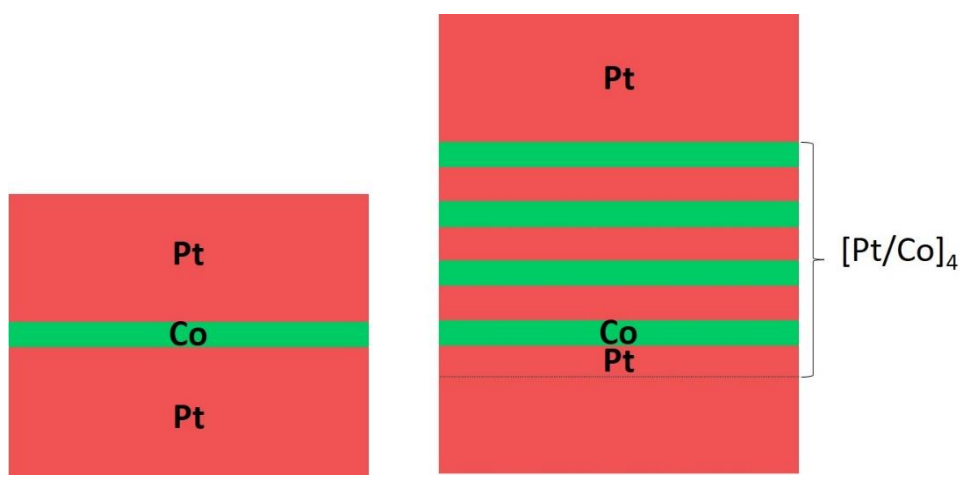

**Figure 1:** Schematic structure of the (left) Pt/Co/Pt trilayer and (right) Pt/[Pt/Co]$_4$/Pt multilayer thin film stacks.

The **magnetic properties** were studied at room temperature by means of a commercial Vector Vibrating Sample Magnetometer (VSM model 10—MicroSense). The magnetization loops were corrected for the diamagnetic contributions of the sample holder and the substrate.

**XMCD and XRMR measurements** were carried out at the BOREAS beamline of the ALBA synchrotron light source [47]. Both samples were maintained at room temperature during these investigations. XAS/XMCD measurements were performed at the HECTOR XMCD vector magnet endstation, using both total electron yield (TEY) and total fluorescence yield (TFY) detection, with the radiation impinging at normal incidence on the sample surface and along the applied field direction. Spectra with right- and left-handed circular x-ray polarization (or positive and negative helicity, respectively) were acquired varying the photon energy across the Co $L_{2,3}$ edge (almost 100 % circularly polarized radiation) and the Pt $M_3$ and $M_2$ edges (70% circular polarization) and applying a magnetic field of up to 5T perpendicular to the film surface, i.e., along the magnetic easy axis of the sample.

XRMR spectra were acquired in the MARES scattering end station in the photon-energy regions corresponding to the Co $L_{2,3}$ and the Pt $M_3$ edges. The intensity of the reflected radiation was measured both by varying the grazing angle $\theta$ at constant photon energy $h\nu$ ($\theta - 2\theta$ scans) and by changing the photon energy at a fixed grazing angle $\theta$ (energy scans). It is worth recalling that the X-ray reflectance signal strongly decreases with the grazing angle $\theta$. In order to maximize the magnetic contribution at grazing incidence, reflectivity measurements were carried out applying a strong in-plane external magnetic field of ±2 T that oriented the magnetic moments of Co and Pt parallel to the sample surface. The reflectance $R(h\nu, \theta)$ at a specific photon energy $h\nu$ and grazing angle $\theta$ is defined as the ratio between the reflected intensity $I_R(h\nu, \theta)$ measured at the same values of $h\nu$ and $\theta$ and the incident intensity $I_0(h\nu)$. The contribution due to the atomic magnetic moments was evaluated acquiring spectra with opposite photon helicity (right or positive, +, and left or negative, −) as well as with opposite in-plane magnetization direction, parallel (↑) or antiparallel (↓) to the X-ray beam direction. The dichroic reflectivity was then calculated as $\Delta R = \frac{1}{2}(R^{\uparrow+} - R^{\uparrow-} - R^{\downarrow+} + R^{\downarrow-})$. The simple average of these spectra, calculated as $\bar{R} = \frac{1}{4}(R^{\uparrow+} + R^{\uparrow-} + R^{\downarrow+} + R^{\downarrow-})$, approximately cancels the magnetic contribution and gives information about the elemental distribution in the investigated structure. The reflectivity spectra, measured under different conditions of helicity and magnetization direction, were obtained as the average of numerous repeated acquisitions that improve the signal-to-noise ratio. For the energy scans across the weak Pt $M_3$ edge at various fixed grazing angles, at least 8 repeated acquisitions lasting ~5 minutes each were necessary.

**Analysis of XRMR measurements** was carried out following a general procedure already described in the literature[48–51]. Averaged and dichroic reflectivity spectra were fitted using the ReMagX code [52], which calculates the reflectance spectra from a magnetic multilayer using the Zak tensor formalism for magneto-optics [53], derived from the general 4×4-tensor formalism for the propagation of light in anisotropic materials [54,55]. In the advanced element-specific simulation mode the dielectric tensor of every layer is calculated from the atomic charge and magnetic scattering factors $f(E) = f_1(E) - if_2(E)$ and $f_m(E) = f_{1m}(E) - if_{2m}(E)$, respectively, where $E = h\nu$ is the photon energy [43]. The determination of the atomic scattering factors for the various elements is the first step in the analysis of the resonant reflectivity spectra [51]. For values of photon energy far from the absorption edges of an element, tabulated data [56] are employed for the charge scattering factor, whereas the magnetic scattering factor is null. In particular, the tabulated data are always used for an element if none of its absorption edges is measured in photon energy scans (Si, O, and Ta in our case). The TEY-XAS spectrum at the absorption edge of an element is assumed as directly proportional to the absorption cross section $\Sigma(E) = 2hcr_0 f_2(E)/E$ in the photon energy range across that edge and, in the same way, the XMCD spectrum is assumed as directly proportional to $f_{2m}(E)/E$. If we subtract to $\Sigma(E)$ a linear and a step-like backgrounds, which represent the contribution of the lower-energy edges and of the non-resonant excitation channels, respectively, its integral on the photon-energy range that includes the threshold is defined as the white-line intensity. The $L_{2,3}$ white-line intensity for 3d transition metal is directly proportional to the number of holes in 3d states and results equal to 24.5 Mb·eV for metallic Co [57]. $\Sigma(E)$ is then conveniently normalized and a suitable background is added to the calculated $f_2(E)$ so that it matches the off-resonance tabulated data. For the Pt $M_3$ edge the white-line intensity is much smaller than the background contribution and $f_2(E)$ is normalized simply matching the values in the pre-edge and post-edge regions with those of the tabulated data. The imaginary part of the magnetic scattering factor $f_{2m}(E)$ is calculated from the XMCD spectra using the same scaling factor used for the XAS spectra. $f_{2m}(E)$ is null far from the absorption-edge region. Finally, the real parts $f_1(E)$ and $f_{1m}(E)$ are calculated through the Kramers-Kronig relations [43,51]. The obtained complex functions $f(E)$ and $f_m(E)$ for Co and Pt in the resonant regions are reported in **Figure S4, Supporting Information**. The same elemental atomic scattering factors are used for the fitting of the reflectivity measurements in the trilayer and the multilayer samples.

The set of average reflectance spectra $\bar{R}$, $\theta - 2\theta$ scans and energy scans, are the first to be fitted considering a non-magnetized model system. The layers in the stack are delimited by an error-function profile [58], characterized by a root-mean-square (rms) roughness $\sigma$. Fitting parameters for the $\bar{R}$ spectra are the thickness and rms roughness of the various layers and the maximum molar concentration of Co. The maximum molar concentrations of the other element, i.e., the value reached sufficiently far from the interfaces, is fixed at the nominal bulk value.

After the structural properties of the stack are determined, the sets of dichroic reflectance spectra $\Delta R$ for the Co $L_{2,3}$ and the Pt $M_3$ edges are separately fitted in order to obtain the Co and Pt magnetization depth profiles. In these final fitting procedures, an in-plane magnetization parallel to the scattering plane is considered for both Co and Pt to replicate the experimental conditions. All the average and dichroic reflectivity spectra and their best fitting curves can be found in the Supporting Information while a selection of them is reported in the next section.

A very detailed recipe for the analysis of XRMR data, followed on general lines in the present work, is exhaustively described by Krieft et al. [51].

**Transmission electron microscopy** (TEM) and **scanning-transmission electron microscopy** (STEM) techniques were used for the micro- and nano-structural characterization. TEM observations were performed using a Philips CM200 microscope operating at 200 kV and equipped with a LaB$_6$ filament, while the analytical high-resolution microscopy was carried out with a probe Cs-corrected FEI Titan[3] G2 60-300

STEM equipped with a X-FEG field emission gun and a Super-X EDX detector system developed at FEI (FEI application note AN002707-2010. Available from: www.fei.com). The latter one allows also for the chemical composition analysis of the layers down to the nanoscale with the energy dispersive X-ray (EDX) microanalysis. Samples for TEM and STEM-EDX analysis were prepared as cross-sections, using the conventional procedure consisting of cutting in slices the sample sandwiched between silicon substrates. Slices were prepared by grinding on abrasive papers and diamond pastes. Disks with a 3 mm diameter were cut from the slices by an ultrasonic cutter. To reduce time of ion milling, in the last step of the mechanical thinning procedure, each 3 mm disk was mechanically thinned in a central area by a Dimple Grinder (Gatan). Final thinning was carried out by an ion beam system (Gatan PIPS) using Ar ions at 5 kV.

**X-ray photoelectron spectroscopy** (XPS) measurements combined with ion sputtering were carried out using bombardments with $Ar^+$ ions that progressively remove portions of the sample until the $SiO_2$ substrate is completely exposed. XPS measurements were acquired on samples as inserted in the vacuum chamber and after every sputtering process, using a monocromatic Al $K_\alpha$ X-ray source (hv=1486.6 eV) and collecting the photoelectrons with a hemispherical electron-energy spectrometer. The sputtering processes were carried out in a vacuum chamber using an Ar pressure of $1\times10^{-5}$ mbar but using different values of i) $Ar^+$ current intensity (ranging from 3 to 8 µA), ii) $Ar^+$ kinetic energy (0.5-3.0 keV) and iii) sputtering time (10-60 min. and only 2 min. for the first sputtering process).

**Spin-polarized Density Functional Theory** (DFT) calculations have been performed within the Plane Augmented Wave (PAW) approach [59,60] and the Perdew-Burke-Ernzerhof (PBE)[61] approximation for the exchange-correlation functional, as implemented in the Vienna Ab-Initio Simulation Package (VASP) [62–64]. We employed supercells with periodic boundary conditions and fcc[111]-stacking, with an in-plane size equal to a (3x4) repetition of the hexagonal Pt(111)- planar unit cell (in-plane lattice constant $a_{Pt}$=2.77 Å, as from bulk calculations), and a total of 12 layers in the vertical direction. Calculations were carried out with a cut-off energy of 400 eV and a (2×2×1) Monkhorst-Pack grid for *k*-points summation. Within each layer, the ratio of Co and Pt atoms, $n_{Co}/n_{Pt}$, was varied in order to simulate different concentration profiles along *z*. Inter- and intra-layer distances were relaxed by minimizing the atomic forces, with convergence criteria of 0.01 eV/Å on the calculated forces. The supercell length $L_z$ was optimized by total energy minimization with respect to $L_z$ variations. Differences of spin-up and spin-down integrated atom-projected spin-polarized desity of states (DOS) gave the atomic magnetic moments per each atom, $m_{Pt}$ and $m_{Co}$, and thereby the DFT Pt- and Co-magnetization profiles of the given atomic arrangements.

## 3. Results and discussion

Structural and microstructural analyses performed by TEM and selected area electron diffraction (SAED) measurements on the reference trilayer indicate that the film is composed of columnar grains having a preferential oriented growth with the {111} lattice planes parallel to the substrate, as expected when a Ta seed-layer is used [65] (**Figure S1, Supporting Information**).

By STEM measurements it is possible to clearly distinguish Co and Pt layers within the sample up to atomic resolution. **Figures 2a and 2b** show two high-angle annular dark field (HAADF) STEM images of the sample, while the bright field (BF) STEM image is displayed in **Figure 2c**. The individual layers are easily identified owing to the different average atomic number of Co and Pt that result in a different contrast in the images. The relative thicknesses are in agreement with the nominal ones, confirming the accuracy of the deposition technique.

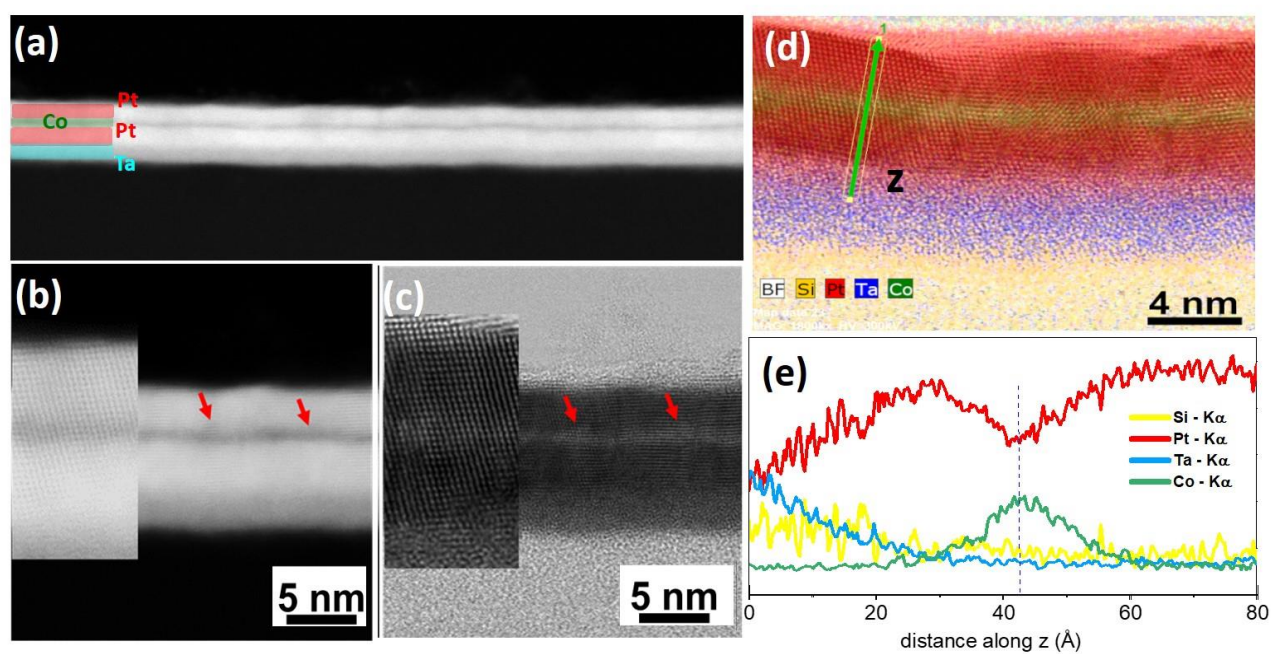

**Figure 2:** Pt/Co/Pt trilayer. (a, b) HAADF-STEM images showing the layered structure of the sample and their epitaxial growth (inset figure b). (c) BF-STEM image corresponding to figure (b); the red arrows evidence a diffuse contrast in the upper Co interface suggesting a larger diffusion of Co atoms in the upper Pt layer; (d) High resolution STEM-EDX map showing the elemental distribution in the sample: Si → yellow, Ta →blue, Pt →red, Co→ green (note that, due to the superposition with the Pt layers, the Co layer appears frequently yellowish); (e) elemental distribution along the arrow averaged over the width of the yellow rectangle marked in (d), revealing an anisotropic diffusion of Co atoms in the Pt layers (larger diffusion in the Pt upper layer, right-side).

**Figures 2b** and **2c** allow to compare HAADF- and BF-STEM images of the same area of the sample, together with the relative high-resolution images in the insets; in the latter, the atomic planes are clearly continuous from the bottom to the top of the sample without deformations inside the Co layer in both images. It is thus possible to state that, even in the BF-STEM image, the visibility of the Co layer is mainly due to the atomic number contrast. Looking at the Co contrast in the images, there is evidence that the (Pt/Co)$_{bottom}$ interface is sharper and more uniform than the (Co/Pt)$_{top}$ one, which is jagged and diffuse (red arrows) thus suggesting a greater diffusion of Co atoms in the upper Pt layer. Such result is confirmed by EDX measurements (**Figure**

**2d**): as above, the (Co/Pt)$_{top}$ interface resolved by EDX appears more diffuse than the (Pt/Co)$_{bottom}$ one, which is definitively sharper. Further quantitative evidence of this asymmetry is given by the compositional analysis performed along the line evidenced in Figure 2d: in the elemental distribution along z (Figure 2e), the minimum of the Pt concentration (red line) corresponds to the Co maximum (green line), whose position is evidenced in Figure 2e by the vertical light-blue dashed line. With respect to this line, the Co distribution is clearly asymmetric, revealing a larger diffusion through the upper Pt layer. It must be stressed that similar results are found even analyzing other areas of the sample. We note here that the Ta concentration is different from zero even inside the top Pt layer, which indicates a partial Ta diffusion within the Pt layer. Such a result has been further confirmed by the XPS analysis (**Figure S12**).

The presence of the observed interdiffusion does not hinder the onset of PMA that is favored by the (111) orientation. Indeed, magnetic measurements show a typical squared hysteresis loop for magnetic field applied perpendicular to the sample's plane (**Figure S2**). The effective saturation magnetization $M_{eff}$, obtained by dividing the measured saturation moment by the nominal Co volume in the film, resulted to be 1700 ±50 emu/cm$^3$ and 1740 ±50 emu/cm$^3$ for the trilayer and the multilayer, respectively. The enhancement of $M_{eff}$ compared to the magnetization of bulk Co (1420 emu/cm$^3$) suggests a significant contribution of the polarized Pt atoms, in line with the data reported in literature [2]. As we shall see, the fitting of the resonant X-ray reflectivity data reveals that, for both the samples, the actual quantity of Co introduced in the films is lower than the nominal one, suggesting that the contribution to the total magnetic moment coming from the magnetized Pt atoms is of the same order of magnitude of that of the Co atoms.

To confirm the presence of a magnetic signal associated to Pt atoms, TEY/XMCD measurements have been carried out on the Pt/Co/Pt trilayer. Normal-incidence TEY spectra measurements acquired on the Co L$_{2,3}$ edge with opposite helicities and their difference, the XMCD signal, are reported in **Figure S3 (a-b)**. Analysis through magneto-optical sum rules [57,66] reveals for a single Co atom a spin magnetic moment of 1.62 Bohr magnetons ($\mu_B$) and an orbital magnetic moment of 0.22 $\mu_B$ and thus a total magnetic moment of 1.84 $\mu_B$ per Co atom. This is slightly larger than the value of total magnetic moment of 1.7 $\mu_B$ [66] found in bulk metallic Co, in particular for the presence of an enhanced orbital momentum, as is the case for low-dimensional Co structure at the interface with Pt layers [67].

TEY/XMCD measurements across the Pt M$_3$ and M$_2$ thresholds are shown in **Figures S3c** and **S3d**, respectively. Distinct XMCD spectra for opposite directions of the out-of-plane applied magnetic field are reported by the green and orange lines. For both the thresholds, the maximum of the XMCD signals is about 1% of the edge jump, due to the small fraction of Pt atoms involved in the proximity effect and to their modest magnetic moment. For a Pt atom at a sharp Pt/Co interface the total magnetic moment was estimated to be about 0.6 $\mu B$ through angle-resolved XMCD measurements [68]. Nevertheless, the inversion of the XMCD signal for opposite external magnetic fields and the change of sign between the M3 and the M2 edges prove the genuineness of the found dichroism. The same change of sign for the Pt M3 dichroic signal with the reversal of the magnetic field is found in the XRMR measurements. The average TEY and XMCD signals used to obtain the charge and magnetic scattering factors $f(E)$ and $f_m(E)$ are calculated combining the spectra obtained with opposite applied magnetic fields (**Figures S3 e-f**). The charge and magnetic atomic scattering factors for Co and Pt used in the fitting of the resonant reflectivity spectra are reported in **Figure S4**. The application of the XAS/XMCD sum rules to the Pt $M_3$ and $M_2$ edges is usually considered tricky, principally for the great uncertainty in the subtraction of the continuum-background component, especially for the $M_2$ edge [69]. Moreover, most of the Pt atoms in the film are expected to be non-magnetic, in particular those in the top portion of the cap layer which mostly contribute to the TEY signal. So, the XMCD measurements on the Pt $M_3$ and $M_2$ edges largely underestimate the induced magnetic moment of the Pt atoms next to the Co layer.

**Figures 3 (a-h)** show a selection of the average reflectance spectra (θ-2θ scans, **(a-b)**, energy scans, **(c-h)**) acquired on the reference trilayer sample together with their best fitting curves. Additional spectra are presented in **Figure S5**. From the simultaneous best fitting procedure, using the Co total concentration and the thickness and roughness of each layer as fitting parameters, the depth profiles of each atomic element in the sample (Si, O, Ta, Pt, Co) were obtained (**Figure 3i).**

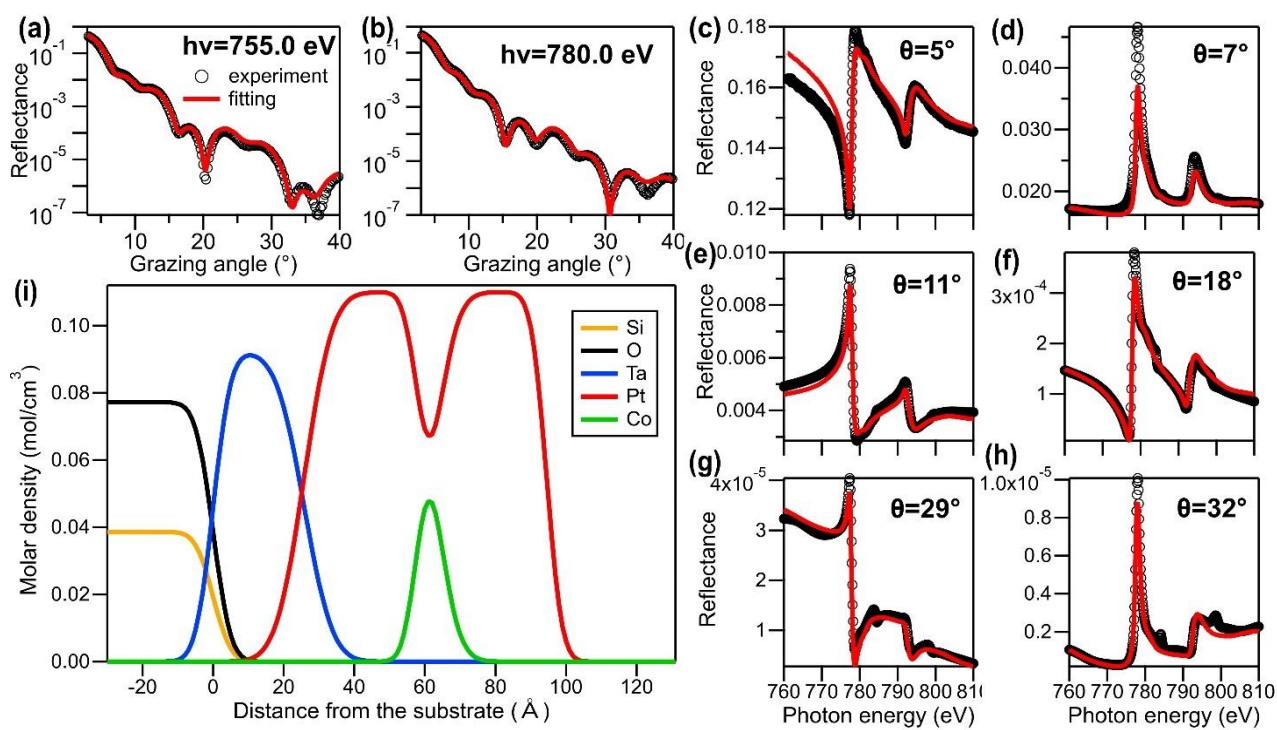

**Figure 3.** Pt/Co/Pt trilayer. (a-h) Measured average reflectivity spectra (empty black dots) and corresponding curves obtained from best fitting parameters (red lines): (a-b) $\theta$-$2\theta$ scan at fixed photon energy (780.0 eV) and (c-h) energy scan on the Co $L_{2,3}$ edge at fixed grazing angle ($\theta$= 5° - 32°). (i) Elemental depth profile as obtained from the fitting of all the angular and energy scans.

As evidenced, the change in the Co/Pt composition profiles is not sharp, confirming that a pure Co film sandwiched between two homogeneous Pt layers was not obtained. Instead, the Co layer shows a broadened atomic distribution with a full width at half maximum (FWHM) of 10 Å inside a thick Pt layer, reaching a maximum atomic percentage of 40%. Integrating the Co depth profile distribution, an areal density of 3.05 x $10^{15}$ atoms/cm$^2$ is obtained, which is much lower than the nominal density of 5.46 x $10^{15}$ atoms/cm$^2$ expected for a pure 6 Å thick Co layer. This discrepancy can be partially ascribed to the uncertainty in the determination of the absolute value of the charge scattering factor for Co, $f_1 - if_2$, which is based on the correlation between the integral of the absorption cross section $\Sigma(E)$, which is directly proportional to $f_2(E)$, and the number of 3d holes per transition-metal atom [57]. Co distribution is also slightly asymmetric with respect to its peak value. The origin of such broadened (with respect to the 6 Å nominal value) and asymmetric measured Co profile may be in principle ascribed to different factors, such as geometrical roughness, interdiffusion between Pt and Co, or different thickness of the Co layer. STEM observations do not support a different thickness of the Co films across the sample, while the presence of an asymmetric interdiffusion or geometrical roughness at the Pt/Co interfaces can be investigated by simulation data. Indeed, following the ReMagX model [52], the shape of the Co distribution is described here with the product of a straight and reversed normal cumulative functions (i.e. scaled error functions), each one characterized by a σ parameter, which corresponds to the root mean square (rms) roughness. According to the fitting procedure, the bottom portion of the Co distribution (corresponding to the (Pt/Co)$_{bottom}$ interface) shows a rms value of σ =3.3 Å (similar to the roughness found for the interface between the substrate and the Ta buffer layer), whereas a broader profile with σ=5.4 Å is found for the (Pt/Co)$_{top}$ interface, thus suggesting a more pronounced Co-Pt intermixing in the topmost part of the Co distribution, closer to the surface.

From this insight on the structural composition of the Pt/Co/Pt trilayer obtained by the XRMR and TEM measurements and analyses, we have built up a computational framework for the DFT theoretical investigation, with the aim to address the interplay between atomic interdiffusion and induced Pt magnetism.

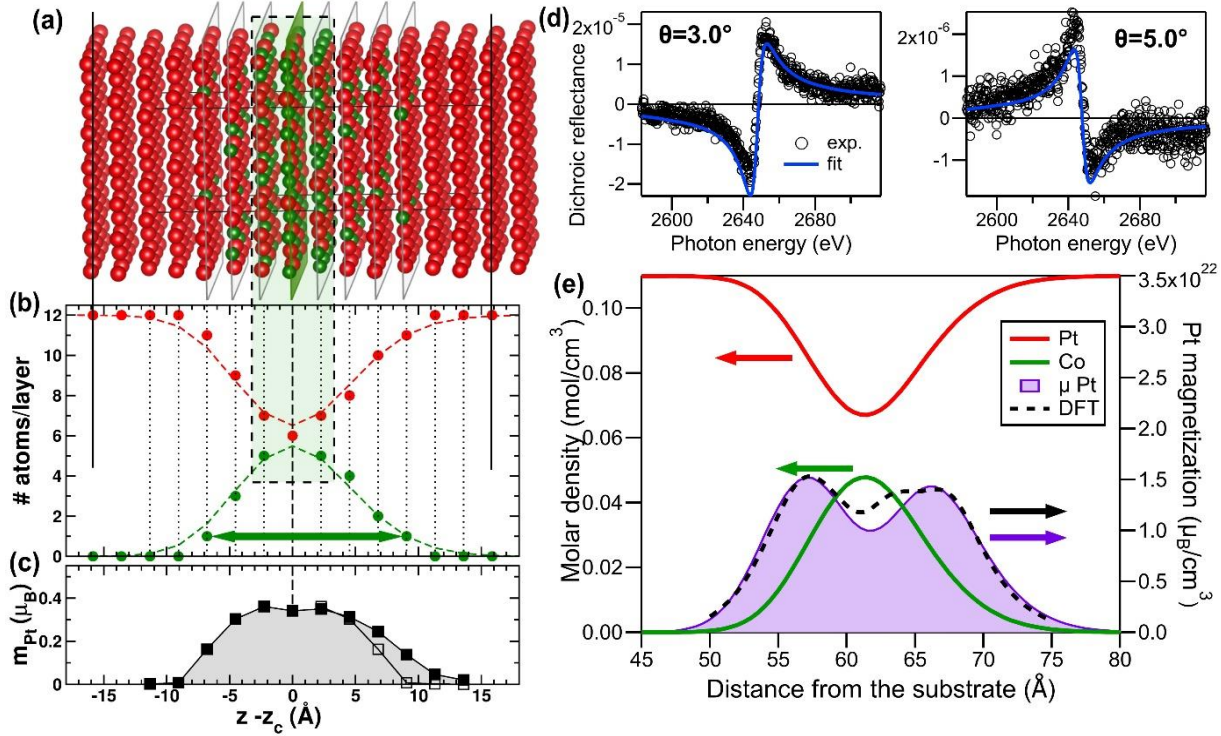

**Figure 4.** Pt/Co/Pt trilayer. (a) Supercell for DFT calculations with Pt (red spheres) and Co (green spheres) atoms. (b) Concentration profiles for Pt (red) and Co (green), as obtained from XRMR depth profiles of figure 3(i) (dashed lines) and as used for the supercell DFT calculations (spheres); (c) Averaged magnetic moment per layer for Pt atoms (black squares, filled). Values calculated considering the ideal case of symmetric interdiffused interfaces are also reported (black empty squares). (d) Measured dichroic reflectivity scans as a function of photon energy across the Pt $M_3$ edge (empty black dots) for $\theta$ =3° and 5° and the best fitting curve (blue lines). (e) Spatial distribution of Pt magnetic moment in the Pt/Co/Pt trilayer as obtained from the fitting of the dichroic energy scans (violet shaded area). The elemental depth profiles of Pt and Co in the region of interest are also reported (solid red and green lines), together with the DFT results for Pt magnetization (dashed black line).

At first, we have calculated the magnetic features for an ideal structure consisting of three layers of Co atoms stacked along the [111] direction and sandwiched between Pt layers. In this case, corresponding to layer-by-layer growth conditions without interdiffusion, concentration and magnetization profiles are symmetric at the top and bottom interfaces: Pt-induced magnetization is larger on atoms within the first interface layer, where $m_{Pt}$ = ~0.3 $\mu_B$, and it decreases rapidly in the two adjacent Pt-layers, being zero within 5 Å from the interface. This agrees well with the value and findings of magnetic-resonant Surface X-ray Diffraction on Co/Pt epitaxial interfaces [70]. Then, asymmetrical interdiffusion between top and bottom interfaces has been taken into account in the calculations by varying the Co and Pt concentration from ideal stepwise behavior, trying to mimic at best the concentration profiles inferred from the experiments. We tested the density profiles obtained by fitting the reflectivity curves, which were qualitatively confirmed by the EDX analysis. **Figure 4a** shows the chosen atomic arrangement that realizes, in the (4x3) supercell geometry, the measured profiles, while **Figure 4b** reports on DFT Pt and Co concentration in each atomic plane of the supercell (filled red and green spheres, respectively), almost perfectly corresponding to the concentration per layers extracted from the XRMR experiments (dashed red and green lines). Black vertical dashed lines mark the position of the atomic layers and the central Co layer ($z=z_c$) is indicated by the thickest line. In both panels, a shaded region shows the nominal Co portion of the trilayer. As deduced from experimental data, the Pt concentration does not go to zero in the part of the trilayer nominally assigned to Co, remaining as large as 50% in that region. Similarly, the Co concentration is different from zero in the first 2-3 atomic layers (~7 Å) away from the nominal Co-Pt interface positions. In this atomic representation, it can be noticed that the asymmetry of measured Co density profile is rather small: Co concentration for $z>z_c$ ((Co/Pt)$_{top}$ interface) decays to zero only one atomic layer further (~2 Å) with respect to its behavior obtained for $z<z_c$, (Pt/Co)$_{bottom}$ interface. Yet, the concentration asymmetry is consistently detected in all XRMR fits. Such small concentration asymmetry induces an asymmetric calculated magnetization profile for Pt, shown in **Figure 4c** (filled squares). The DFT Pt-magnetization profile plotted here was obtained as the sum of all Pt-atomic magnetic moments $m_{Pt}$ in a given layer divided by the total number of Pt atoms in the layer. Due to the still relatively large (50%) Pt concentration in the Co-part of the trilayer, Pt-magnetization density does not vanish and its profile only shows a small valley. Moreover, in the top region with larger Co-Pt intermixing, Pt magnetization extends up to ~10 Å from the interface, while a more confined magnetization region (∼5 Å thickness) shows up in the bottom region with fewer intermixed layers. The magnetization asymmetry can be better appreciated when comparing the results with those obtained for the case of symmetric interfaces: **Figure 4c** reveals the symmetric Pt-magnetization profile (empty squares) obtained for a supercell with the same Pt and Co concentration profiles for $z > z_c$ and $z < z_c$ (realized with different atomic configurations on the left and right regions). Integrating the calculated depth profile distribution of Pt magnetic moment and assuming for the Pt/Co/Pt stack, an atomic density equal to that of pure Pt, we obtain for the trilayer sample a Pt magnetic moment per unit area of 2.43 x 10$^{15}$ $\mu_B$/cm$^2$. Considering a uniform magnetization of the Co atoms with 1.84 $\mu_B$ per atom, the areal density of Co magnetic moment results in 6.00 x 10$^{15}$ $\mu_B$/cm$^2$, leading to a total magnetic moment per unit area of 8.43 x 10$^{15}$ $\mu_B$/cm$^2$. This must be compared with the value of (11.1±0.5) x 10$^{15}$ $\mu_B$/cm$^2$ provided by the VSM measurements. The relatively small discrepancy between the XRMR/DFT analysis and the macroscopic measurements can be attributed to a slight underestimation of the absolute quantity of Co in the fitting of the resonant reflectivity spectra, as stated above.

Finally, we compare the DFT calculations with the XRMR results for Pt-induced magnetization. The dichroic reflectivity spectra across the Co L$_{2,3}$ can be satisfactorily reproduced with a distribution of Co magnetic moments that exactly matches the Co chemical depth profile (see **Figure S6**). **Figure 4e** shows the concentration depth profile of Pt magnetic moments as obtained by the best fitting of the reflectance energy scans on the Pt M$_3$ edge, together with Pt (red line) and Co (green line) density distribution. Two examples of the reflectivity energy scan (empty circles) at $\theta$=3.0° and 5.0° and their best fitting curves (blue lines) are reported in **Figure 4d**. Additional spectra at all the considered grazing angles are reported in **Figure S7**. The distributions of Pt magnetic moments provided by the fitting of XRMR measurements is characterized by a double-peak structure with a narrow valley at the center corresponding to the minimum in the concentration of Pt atoms. The distribution of Pt magnetic moments has a FWHM (referred to the height of the two peaks) of 16 Å and, as expected, it is broader than the distribution of Co atoms which induce proximity magnetization in Pt. The penetration of induced Pt-magnetization is larger at the (Co/Pt)$_{top}$ interface, confirming the above observed asymmetry. Its quantitative estimation depends on the definition of the top and bottom region, which is not trivial here, due to the low nominal Co content and its strong intermixing. Nevertheless, we have calculated the ratio of the integrals of Pt magnetization in the two regions (details in **Figure S8**) and obtain a value of top/bottom equal to 1.25. These features are very similar to those of the calculated DFT Pt-magnetization profile discussed above, as it can be appreciated by superimposing the DFT magnetization profile with the profile obtained by the fitting of the XRMR measurements in **Figure 4e**. The almost perfect agreement gives us confidence both in the robustness of the structural analysis obtained by TEM and XRMR, as well as in the validity of ascribing the asymmetric proximity magnetization profile to the different Co distribution and interdiffusion at the two interfaces.

In a further step, this detailed investigation has been extended to a complete multilayer with four bilayer repetitions (N=4).

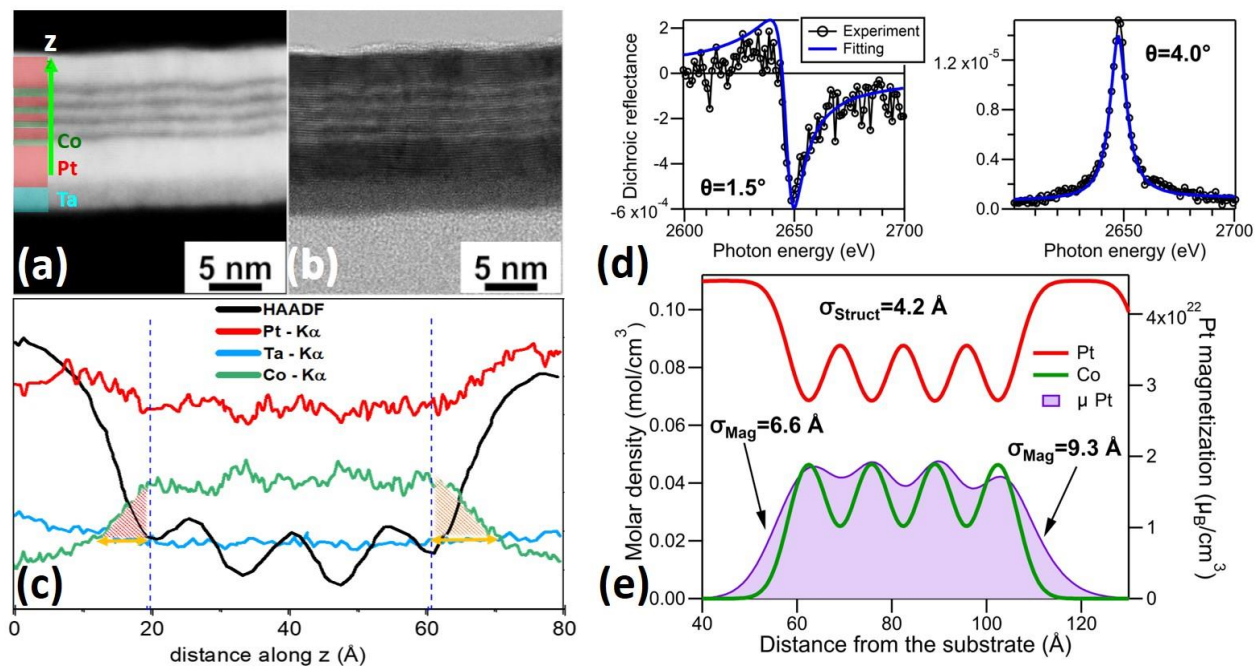

**Figure 5.** [Co/Pt]$_4$ multilayer. (a, b) HAADF and corresponding BF-STEM images. The contrast is sensitive to the atomic number allowing to distinguish the Co, Pt, Ta layers and the SiO$_x$ substrate. (c) Elemental STEM-EDX analysis performed along the arrow shown in (a); (d) Experimental XRMR spectra (black empty circles and lines) and best fitting curves (blue lines). (e) Distribution of Pt magnetic moments (violet-shaded area) compared with Pt (red line) and Co (green line) concentration depth profiles.

STEM studies performed on the multilayer sample have confirmed the realization of the designed structure. In particular, **Figure 5a and 5b** show a typical HAADF- and the corresponding BF-STEM image of the sample. The contrast variations allow the different layers to be identified, although a certain degree of intermixing can be inferred among the Co and Pt layers. EDX maps and compositional analysis performed along selected lines have barely revealed the presence of the different layers. **Figure 5c** shows the EDX atomic concentrations of the multilayer along the line profile marked in **Figure 5a**. The Co and Pt concentrations are indicated by the red and green lines, respectively. To evidence their small variations, the dark line reproducing the contrast variation in the HAADF image is also included. It is very difficult to see variations of the Co and Pt concentrations inside the multilayer, while they are better defined at the $(Pt/Co)_{lowest}$ and $(Co/Pt)_{upmost}$ interfaces. The dashed blue lines show the position of the lowest and uppermost Co layers. Looking at the Co curve at the edges of the multilayer (two shaded regions), it is possible to observe a certain tendency of the Co atoms to diffuse more in the $(Co/Pt)_{upmost}$ interface than in the $(Pt/Co)_{lowest}$ one, reproducing for the edges of the multilayer the same behavior observed in the trilayer. It is important to underline that the asymmetric diffusion of the inner Co and Pt layers is beyond the resolution of this technique.

**Figure 5d** shows two examples of XRMR scans at the Pt $M_3$ edge (fixed grazing angle $\theta$=1.5° and 4.0°) with their best fitting curves. The spectra obtained for other values of $\theta$ are shown in **Figure S11**. **Figure 5e** reports the Pt magnetization depth profile (violet dashed area) obtained by the fitting of the XRMR spectra superimposed to the Pt and Co concentration distribution provided by the fitting of the average reflectivity scans (**Figure S9**). As in the case of the trilayer sample, we did not observe distinct Co layers but a modulation of the Co concentration inside the Pt layer with four peaks of the chemical depth profile. The distribution of Pt magnetic moments (violet area) obtained by the analysis reflects the chemical distribution of Co. A detailed magnetization profile with four valleys coincident with the minima in Pt concentration would be much more complex than that found in the trilayer sample and is beyond the resolution of XRMR. As in the case of the trilayer, the XRMR analysis confirms a larger penetration (in terms of spatial extent) of the induced magnetization towards the surface, the magnetic Pt distribution at the $(Co/Pt)_{upmost}$ interface having a rms roughness $\sigma$ =9.3 Å, whereas for the lowest interface the fitting provides a rms roughness of 6.6 Å. Moreover, a weakening of the induced magnetic moment (about -10%) is evidenced in correspondence of the upmost Co layer. As in the case of the trilayer sample, the distribution of Co magnetic moments perfectly matches the chemical distribution as revealed by the fitting of XRMR across the Co $L_{2,3}$ edge (see **Figure S10**).

The possibility of extending the experimental approaches for studying the asymmetric effect in trilayer to a multilayered system allowed ruling out the influence of the substrate/buffer layer and to correlate the observed behavior with the specific growth mechanism characteristic for the two elements. Indeed, although the interface sharpness is generally expected to decrease with thickness and/or number of layers, a combination of thermodynamic and kinetic factors, strictly related to the growing mechanism and the conditions applied during the fabrication process, may lead to different results, as in our case. In particular, during the sputtering deposition, different intermixing processes may occur, even at room temperature, between the impinging and surface atoms, thus resulting in wide interfaces and significant roughness. A prediction of the resulting quality of the interface in case of two different transition metals has been recently given by Chandrasekaran et al. [71]. The authors provided a semi-empirical model mainly based on the surface energy and the crystal structure properties of the metals involved in the growth, which were demonstrated to play a key role in determining the intermixing events occurring during a sputtering deposition. According to their model, two main mixing processes are expected to take place over the layer surface during the growth: i) a deposition-induced exchange due to the ballistic collision between incident and surface atoms ( that depends on several factors such as the energy of the incident film atom, the atomic mass, interatomic distance, etc) and ii) a thermodynamic exchange of atoms between the surface and the subsurface layers driven by the reduction in the surface-free energy of the system. When the deposition is carried out at room temperature, they assumed that such exchange processes are limited to the surface and the subsurface atoms; in other words, diffusion of atoms in the bulk as well as desorption atoms from the surface, which are both temperature-dependent, may be considered negligible. Depending on the nature of the two metals, on their atomic radius and their lattice characteristics, it is possible to obtain a different type of interface profile and, most important for us, a different interface roughness is expected, depending on which of them is considered as "substrate" or "growing" element (i.e., Pt growing over a Co substrate or Co growing over a Pt substrate). Thus, according to the model developed in Chandrasekaran et al. [71], it is possible to predict the *effective interface width* ($\sigma$) as

$$\sigma = \frac{1}{0,59} A * \left[1 + e^{-B(\gamma_s - \gamma_g)}\right] \qquad \text{eq (1)}$$

where $\gamma_s$ and $\gamma_g$ are the surface energy of the substrate and the growing metals, respectively, and A and B are two effective parameters accounting for the exchange mechanisms (ballistic and thermodynamically driven) that have been experimentally demonstrated to be mainly dependent on the crystal structure of the substrate and the growing layer [71]. In the case of the Co/Pt interface, the surface energy $\gamma$ is 1.27 eV/atom and 1.51 eV/atom for Co (hcp) and Pt (fcc) respectively, while the A and B parameters, experimentally derived in Chandrasekaran et al. [71], are A= 0.27 ± 0.02 nm; B= 1.3 ± 0.5 eV$^{-1}$ for hcp-fcc interfacing lattices. According to eq. (1), for the Pt/Co/Pt system, $\sigma_{bottom} < \sigma_{top}$ being $\sigma_{bottom} = 0.8\ nm$ and $\sigma_{top} = 1.1\ nm$ ), thus a higher number of intermixing events are expected when the Pt grows onto a Co surface, resulting in a wider interface. The $\sigma$ values calculated from the semi-empirical model are slightly higher than those experimentally found by XRMR and TEM analysis. However, the relative $\sigma$ values provide an explanation for the difference in the intermixing between the bottom and the top interface, and, consequently, for the magnetic depth profile occurring in our samples. On the other hand, our results confirm the applicability of the model to a multilayered system, since the effect of a different intermixing has been experimentally observed even on the last interface of the [Co/Pt]$_4$ multilayer, albeit the presence of a Ta buffer layer and the high number of layers does not allow for a good agreement between the calculated and the experimental data.

## 4. Conclusions

In the present work we have confirmed the occurrence of an asymmetric magnetic proximity effect in the Pt/Co/Pt trilayer, namely a larger spatial extent of the total magnetic moment induced on the heavy metal at the top interface. In order to better elucidate the correlation between the structural and the microstructural feature of the interfaces with the magnetic proximity effect and the consequent induced moment on the heavy metal, a comprehensive experimental and theoretical study has been carried out on a nominally symmetric Pt/Co/Pt trilayer and on a [Co/Pt]$_4$ multilayer. The asymmetry in the induced moment observed in the reference trilayer is preserved even in the multilayer, with a larger extent of the Pt magnetic moment in the (Co/Pt)$_{top}$ interface. The combination of advanced microscopy and spectroscopic approaches supported by DFT calculations allowed the compositional features of the interfaces to be strictly correlated with the induced magnetic moment of the Pt. In particular, the higher interdiffusion of the Co atoms in the (Co/Pt)$_{top}$ and (Co/Pt)$_{upmost}$ interface for the trilayer and the multilayer, respectively, results in a larger extension of the Pt induced magnetic moment in both cases. The sputtering growth mechanism, which brings to different interfaces depending on the deposition order (Co over Pt or vice versa) which is related to the surface energy and to the crystal structure of the two materials involved in the growth, is assumed to play a key role in determining such asymmetry. In other words, the magnetic proximity effect was here intimately correlated to the chemical and structural characteristic of the two metals and to the consequent intermixing events occurring during their growth. Since the interdiffusion processes may be further influenced by the deposition parameters (temperature, pressure, post-deposition treatments), once selected the two materials to be interfaced, the choice of the fabrication conditions may represent a potential way for a fine control of the MPE and the related phenomenology, thus allowing a suitable design of the material system with a better performance in its specific technological application.

## Supporting Information

Additional bright-field TEM images and electron diffraction on the trilayer sample; VSM hysteresis loops on trilayer and multilayer sample; TEY XAS and XMCD measurements; atomic scattering factors calculated from XAS and XMCD measurements; Co L$_{2,3}$ average reflectance spectra on the trilayer sample; Co L$_{2,3}$ dichroic reflectance spectra on the trilayer sample; Pt M$_3$ dichroic reflectance spectra on the trilayer sample; analysis of Pt magnetization profile on the trilayer sample; Co $L_{2,3}$ average reflectance spectra on the multilayer sample; Co $L_{2,3}$ dichroic reflectance spectra on the multilayer sample; Pt $M_3$ reflectance spectra on the multilayer sample; XPS/sputtering depth profile on the trilayer sample.

# Acknowledgements

AV, FO, and AR are grateful to Prof. Giovanni Stefani, Università degli Studi Roma Tre and CNR-ISM, for his constant encouragement and advice. Support to the photoemission measurements by Gianluca Di Filippo and Raffaella Baldassarre is particularly appreciated. AV and FO are thankful to Francesco Borgatti, CNR-ISMN Bologna, for fruitful discussions. AV is grateful to Sebastian Macke, QAware GmbH, for profitable discussions about the potentiality and use of the ReMagX software.

Partial financial support by Università degli Studi Roma Tre, "Piano Straordinario della Ricerca 2015, azione n. 3 -Potenziamento dei laboratori di ricerca- " is greatly acknowledged.

# SUPPORTING INFORMATION

# Disclosing the nature of asymmetric interface magnetism in Co/Pt multilayers

## Figure S1

A typical TEM bright field image of the sample is shown in panel **(a)**. The amorphous Ta layer is clearly visible at the bottom of the sample over which the Pt and Co layers, not separately distinguishable in the image, give rise to dark and bright contrast regions extending from the bottom to the top of the film and recognizable as columnar grains. The selected area electron diffraction (SAED) pattern of panel **(b)** corresponds to the sample imaged in panel **(a)**. The most intense diffraction spots due to the film are indexed as Pt {111} and are aligned with the Si {002} ones of the substrate. Thus, the grains have a preferential oriented growth with the {111} lattice planes parallel to the substrate.

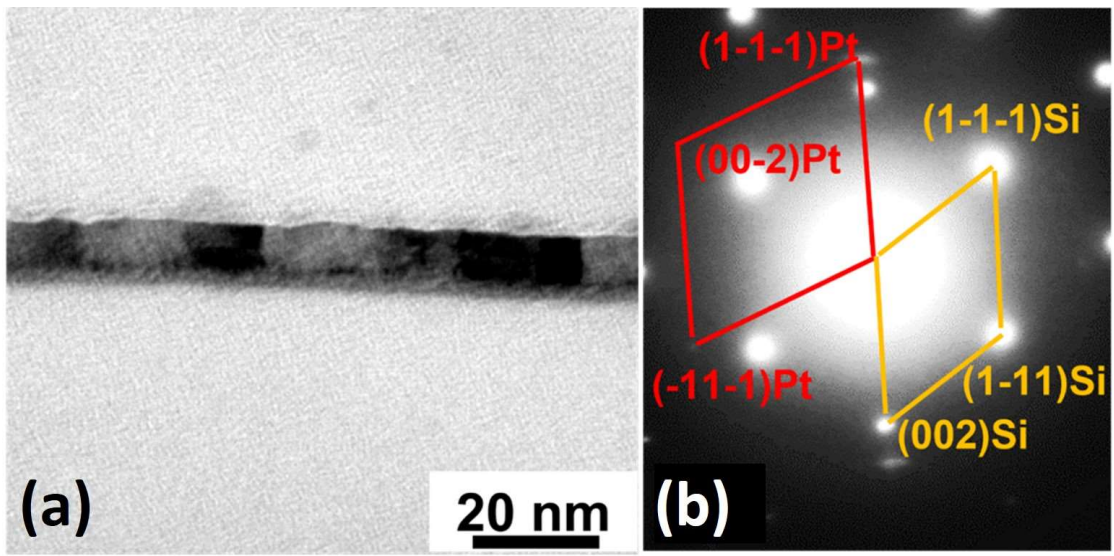

# Figure S2

The graph shows the room temperature hysteresis loop measured through a vibrating sample magnetometer with the magnetic field applied along the film normal for the reference trilayer sample (red markers) and the multilayer sample (black markers).

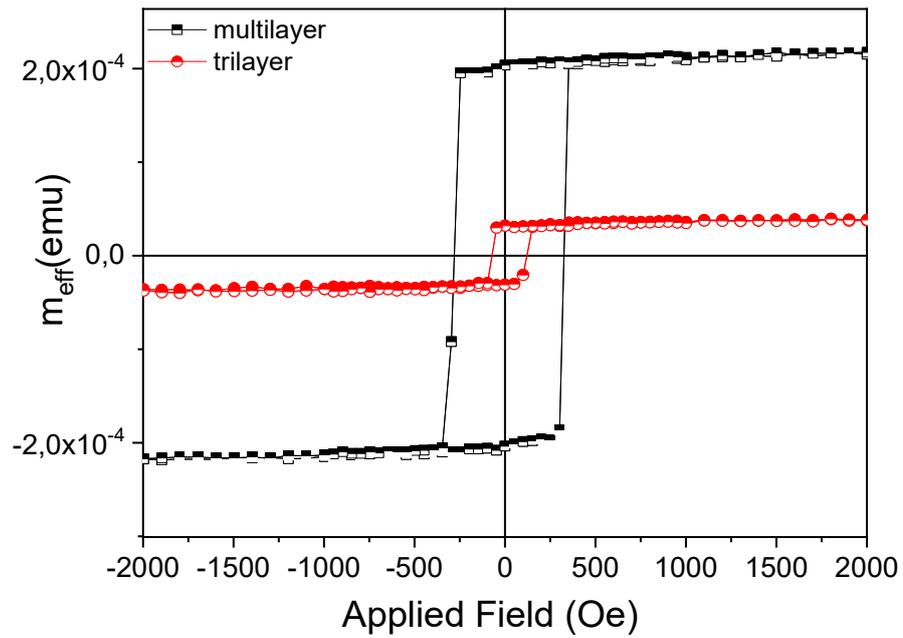

# Figure S3

In the presented data, a TEY spectrum is defined as the current $I$ of photoelectrons emitted from the sample as a function of the photon energy $E = h\nu$, normalized by the current of photoelectrons emitted from a reference gold mesh.

The panels of this figure report TEY/XMCD measurements carried out at normal incidence ($\theta$=90°) and with an external applied magnetic field $\mu_0 H$= 5.0 T perpendicular to the sample surface.

In **panels (a)** and **(b)** measurements across the Co $L_{2,3}$ edge on the trilayer reference sample are shown. **(a)**: TEY absorption spectra across the Co $L_{2,3}$ edge with the magnetic field directed parallel ($\uparrow$) to the wavevector of the incoming radiation for right- (positive helicity $I^{\uparrow+}$, blue line) and left- (negative helicity $I^{\uparrow-}$, red line) 100 % circularly polarized radiation. The black line is the average of the two measured spectra, i.e., the TEY signal expected for a non-magnetized sample. **(b)**: XMCD signal across the Co $L_{2,3}$ edge defined as the difference between the absorption spectra with opposite helicity $I^{\uparrow-} - I^{\uparrow+}$.

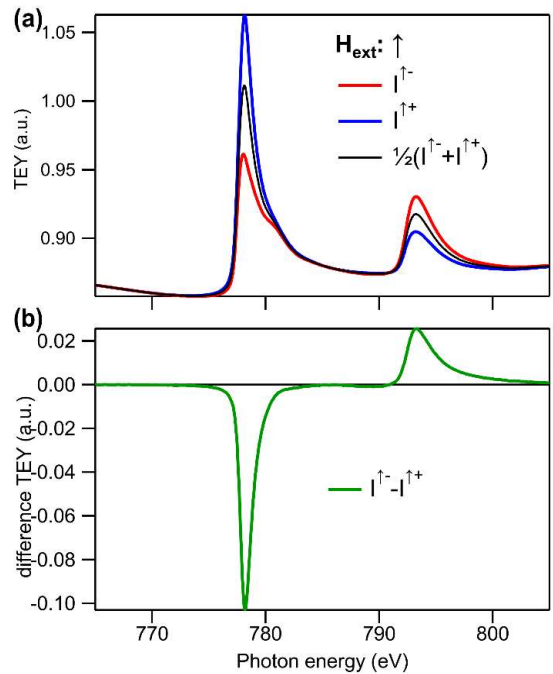

Panels **(c)** and **(d)** report measurements on an additional trilayer sample that differs from the reference sample presented in the paper only in having a thinner Pt cap layer, 15 Å instead of 30 Å. Therefore, its stack sequence is Si(100)/SiO$_x$/Ta(30 Å)/Pt (30 Å)/Co (6 Å)/Pt (15 Å). TEY measurements were caried out across the Pt $M_3$ **(c)** and the Pt $M_2$ **(d)** absorption edges, by varying both the orientation of the magnetic field and the photon helicity (70% circular polarization). In both panels, the black line is the average spectrum defined as $\frac{1}{4}(I^{\downarrow+} + I^{\downarrow-} + I^{\uparrow+} + I^{\uparrow-})$. The green and orange lines are the XMCD spectra for the external magnetic field oriented antiparallel ($\downarrow$) and parallel ($\uparrow$) to the radiation wavevector, respectively. XMCD spectra are multiplied by a factor 20 for a better readability. The inversion of the XMCD signal for opposite external magnetic fields and the change of sign between the $M_3$ and the $M_2$ edges prove the genuineness of the extremely small dichroic effect.

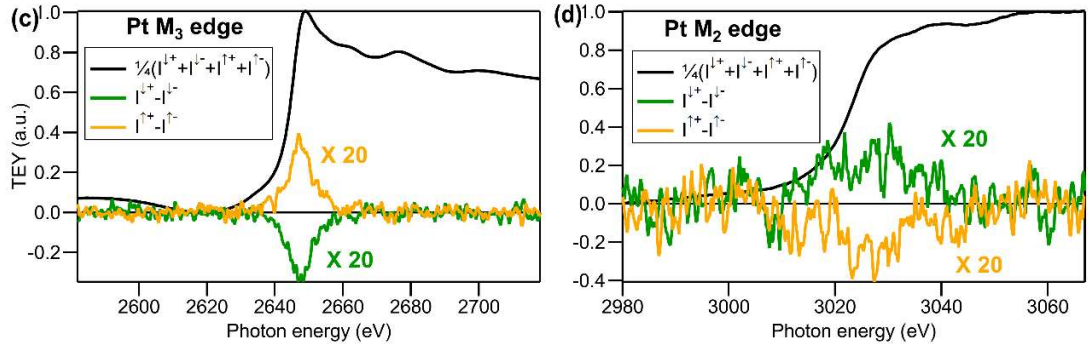

Panels **(e)** and **(f)** show the spectra acquired across the Pt M$_3$ edge on the reference trilayer sample. **(e)**: TEY absorption measurements across the Pt M$_3$ edge with the magnetic field parallel (↑) and antiparallel (↓) to the wavevector of the incoming radiation and with opposite photon helicities (70% circular polarization). The positive spectrum (red line) is the average of the TEY measurements with antiparallel magnetic moment and photon helicity, $\frac{1}{2}(I^{\uparrow -} + I^{\downarrow +})$, and the negative spectrum (blue line) is the average of the TEY measurements with parallel magnetic moment and photon helicity, $\frac{1}{2}(I^{\uparrow +} + I^{\downarrow -})$. The inset better illustrates the contrast between the two spectra at the Pt M$_3$ maximum. The black line is the average of positive and negative spectra, i.e. the TEY signal expected for a non-magnetized sample. **(f)**: XMCD signal across the Pt M$_3$ edge defined as the difference between the positive and negative spectrum.

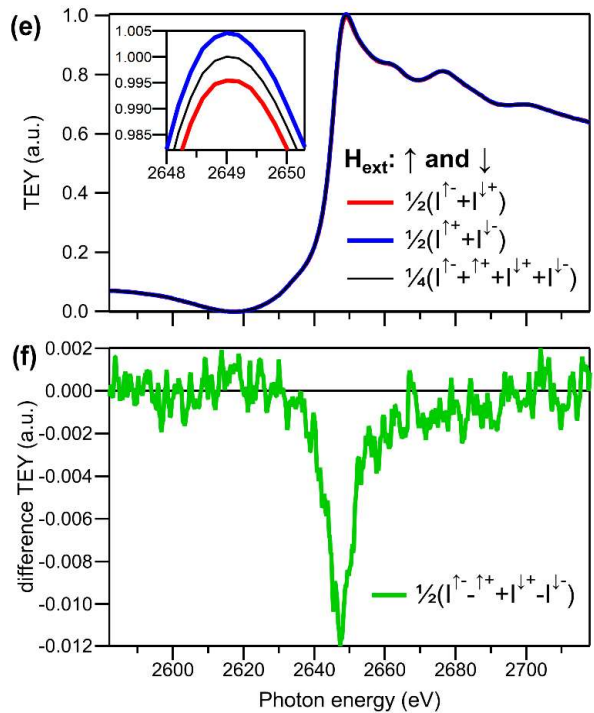

Measurements with lower applied magnetic field (0.1 T) gave almost identical results, the easy magnetization axis being perpendicular to the sample surface.

# Figure S4

In this figure the Co and Pt charge and magnetic atomic scattering factors calculated from the average TEY spectra and the XMCD spectra of **Figure S3** are reported.

Panels **(a)** and **(b)**: charge scattering factor $f(E) = f_1(E) - f_2(E)$ determined from the experimental measurements and Kramers-Kronig relations for Co in the photon energy region around the $L_{2,3}$ edge **(a)** and for Pt in the photon energy region around the $M_3$ edge **(b)**. In both panels, the solid red and black curves represent the real $f_1(E)$ and imaginary $f_2(E)$ parts of the charge scattering factor, respectively. They are compared with the real part (dash-dot red curves) and imaginary part (dash-dot black curves) of the calculated atomic scattering factors in the database by Chantler [J. Phys. Chem. Ref. Data, 29 (2000), 597].

Panels **(c)** and **(d)**: magnetic atomic scattering factor $f_m(E) = f_{1m}(E) - f_{2m}(E)$ for Co around the $L_{2,3}$ edge **(c)** and for Pt around the $M_3$ edge **(d)**. In both panels, the solid orange and blue lines represent the real $f_{1m}(E)$ and imaginary $f_{2m}(E)$ parts, respectively.

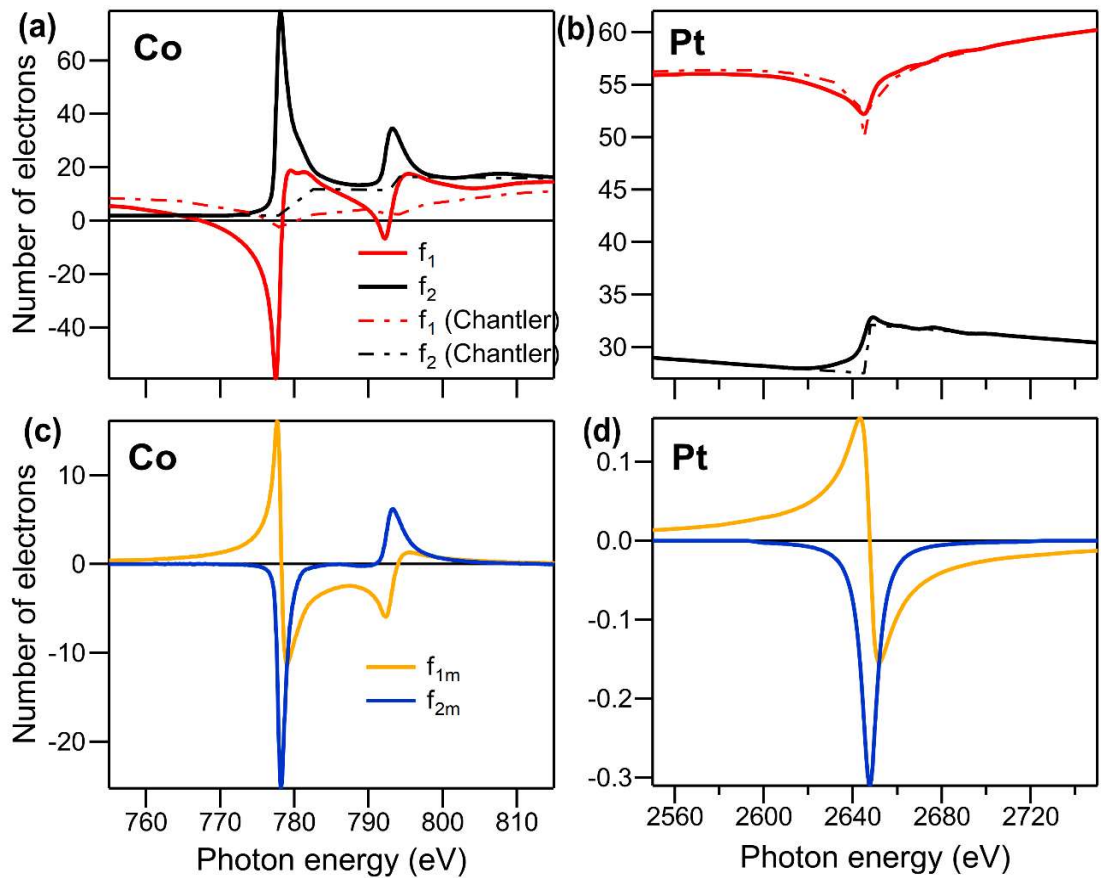

# Figure S5

Measured average reflectance $\bar{R}$ scans in the Co L$_{2,3}$ region on the reference trilayer sample (empty black circles) and the corresponding best fitting reflectivity curves (red lines) that determined the elemental depth profile in **Figure 3i**. Top panels: $\theta - 2\theta$ scans at six values of fixed photon energy (755-815 eV). Bottom panels: photon energy scans at twelve fixed values of grazing angle $\theta$ (5°-36°).

In the energy scans, in particular for larger grazing angles, two small additional peaks around 784 eV and 798 eV are evident but they were not visible in the more surface-sensitive TEY measurements. These peaks can be ascribed to the presence of a Ba contamination at the interface between the SiO$_2$ substrate and the Ta buffer layer, as evidenced by the X-ray photoemission spectroscopy measurements (XPS) described below. These small peaks are not taken into account in the fitting procedure.

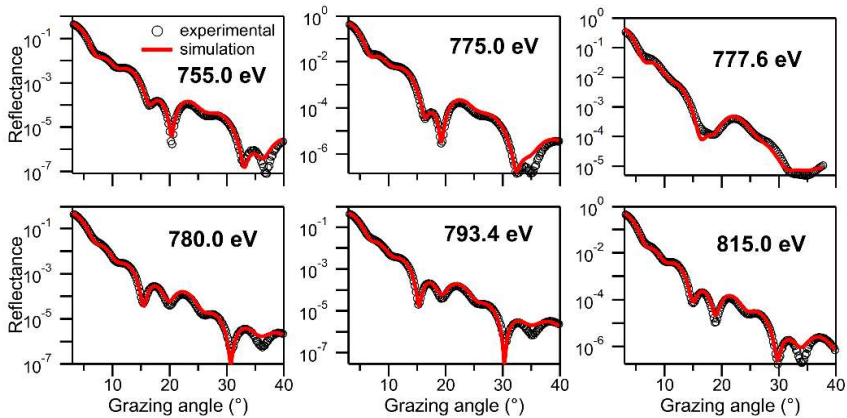

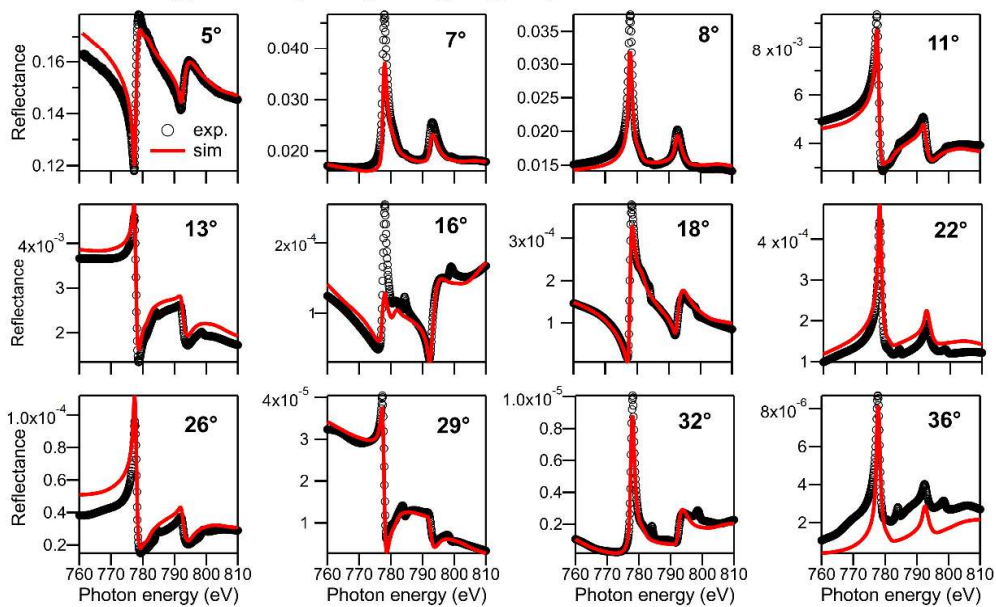

# Figure S6

Dichroic reflectivity spectra acquired on the reference trilayer sample (empty black circles) and the simulated curves (blue lines) that reproduce a uniformly magnetized Co layer, i.e., a Co magnetization profile equal to the Co density profile multiplied by a constant. In the top panels: the asymmetry ratio $AR = (R^{\uparrow+} - R^{\uparrow-})/(R^{\uparrow+} + R^{\uparrow-})$ acquired in $\theta - 2\theta$ scans at fixed photon energy (775.0-793.4 eV). In the middle panels: the dichroic reflectivity $\Delta R = R^{\uparrow+} - R^{\uparrow-}$ acquired in the photon-energy scans at fixed grazing angle (5°-36°). In the lowest panel: the magnetization depth profile for Co superimposed to the elemental depth profile in the trilayer sample.

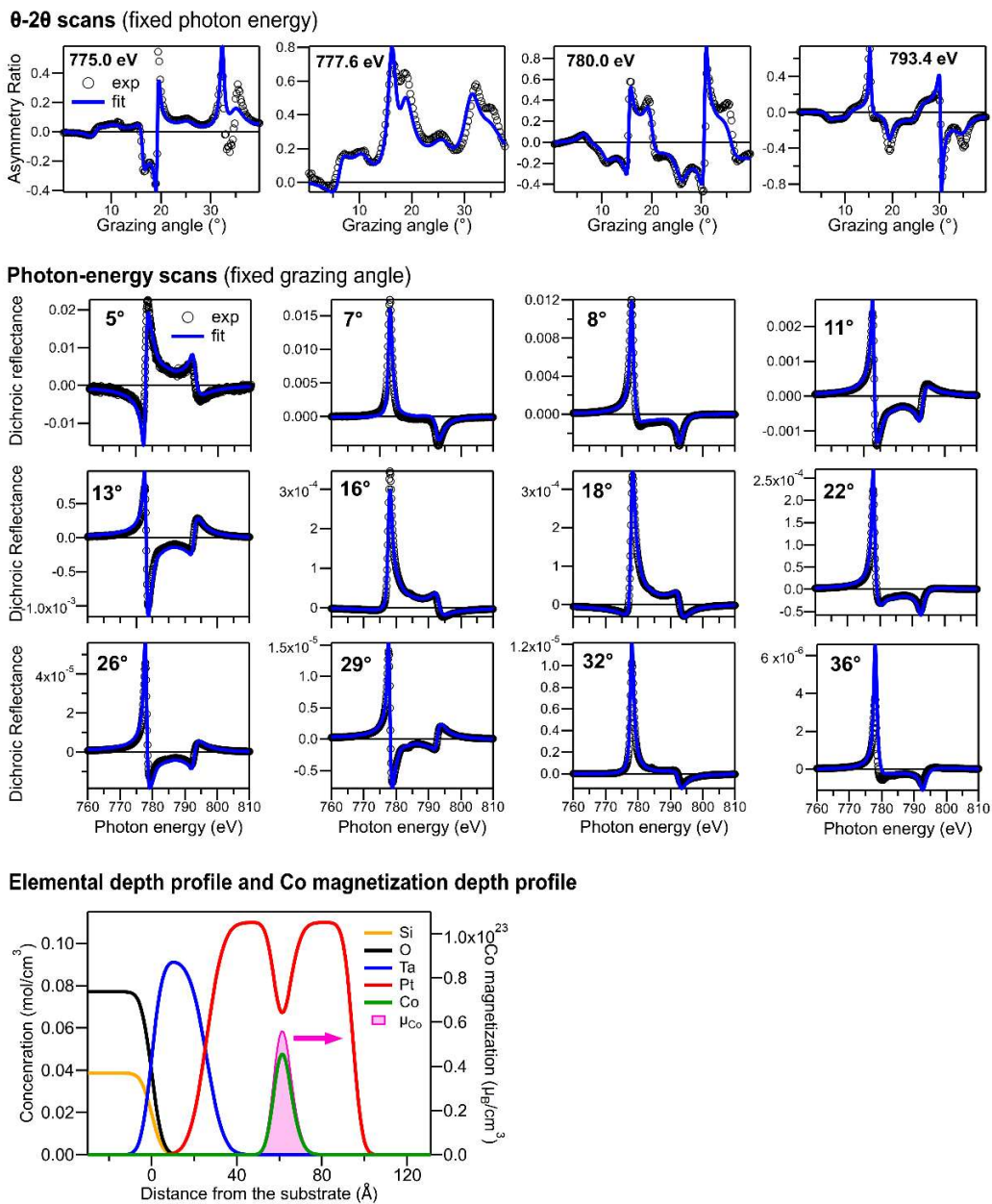

# Figure S7

Measured dichroic reflectivity $\Delta R = \frac{1}{2}(R^{\uparrow+} - R^{\uparrow-} - R^{\downarrow+} + R^{\downarrow-})$ spectra across the Pt L$_3$ edge on the reference trilayer sample (empty black circles) and the best-fitting curves (blue lines) relative to the distribution of Pt magnetic moments reported in **Figure 4e**. Measurements consist of six photon energy scans at fixed grazing angles (3°-5°).

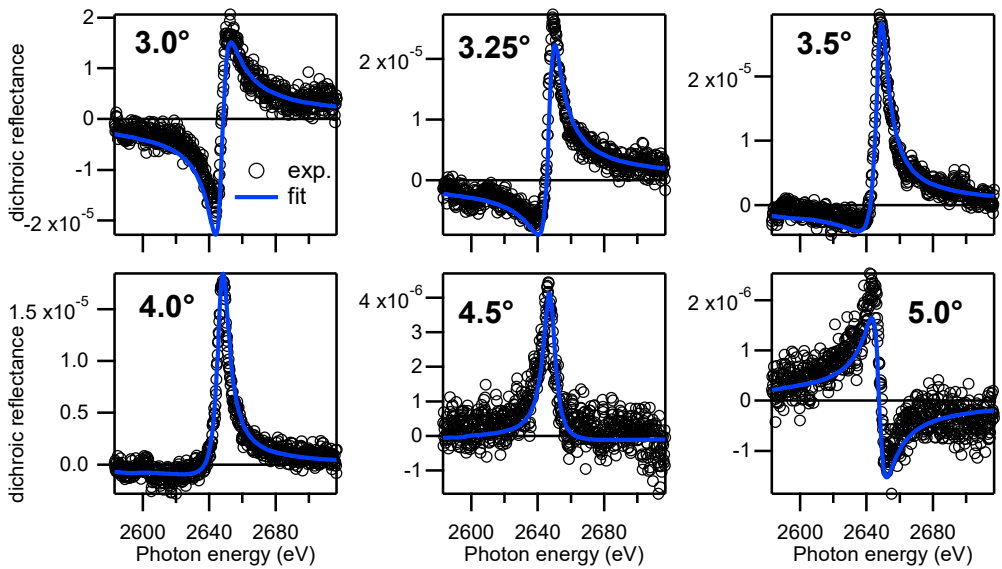

# Figure S8

The graph shows the Co concentration depth profile (green line, left axis) superimposed to the Pt magnetization depth profile (violet line, right axis). Both curves are taken from **Figure 4e**. The depth values $z_1$=56.5 Å and $z_2$=66.5 Å correspond to the half maximum of Co concentration and identify the full width at half maximum (FWHM) of the Co depth profile curve, FWHM=$z_2 - z_1$=10 Å. These two positions can be viewed as the limits of the Co-rich layer. In order to investigate a possible asymmetry in the MPE between the (Pt/Co)$_{bottom}$ and the (Co/Pt)$_{top}$ interfaces, we calculate the integral of the Pt magnetization depth profile in the two regions external to the Co-rich layer, as evidenced by the two shaded areas in the graph. The integral in the bottom region (blue shaded area) corresponds to a magnetic moment per unit area of 4.5×10$^{14}$ $\mu_B$/cm$^2$ while the integral in the top region (red shaded area) is 5.6×10$^{14}$ $\mu_B$/cm$^2$. The induced magnetization in the top region results 25% larger than that in the bottom region.

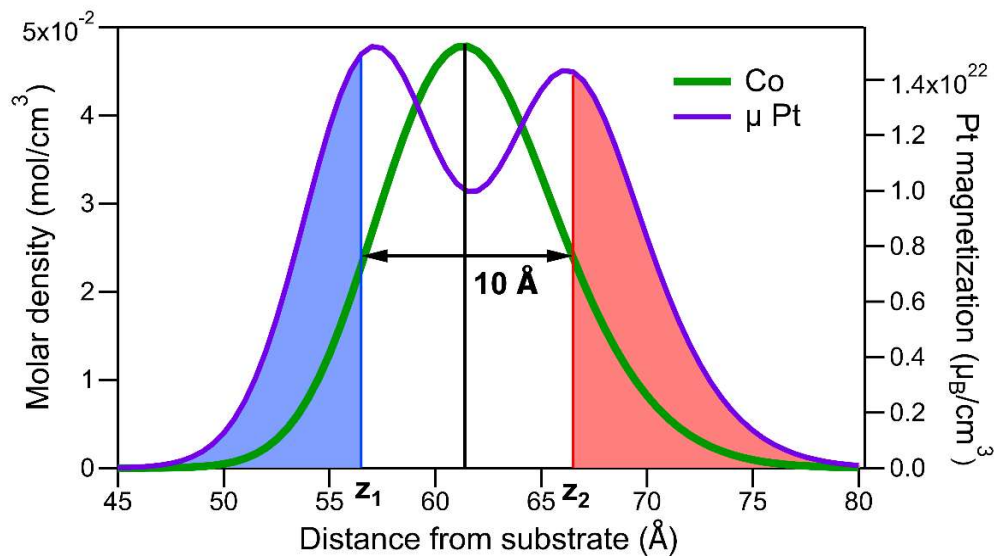
9

# Figure S9

Average reflectance $\bar{R}$ scans in the energy region of the Co L$_{2,3}$ edge measured on the multilayer sample (empty black circles) and the best-fitting curves (red lines) relative to the elemental depth profile presented in **Figure 5e**. Top panels: $\theta - 2\theta$ scans at seven fixed values of photon energy (755-815 eV). Bottom panels: photon energy scans at ten fixed values of grazing angle $\theta$ (5°-38°). The fixed values of photon energy and grazing angle are indicated in the legenda.

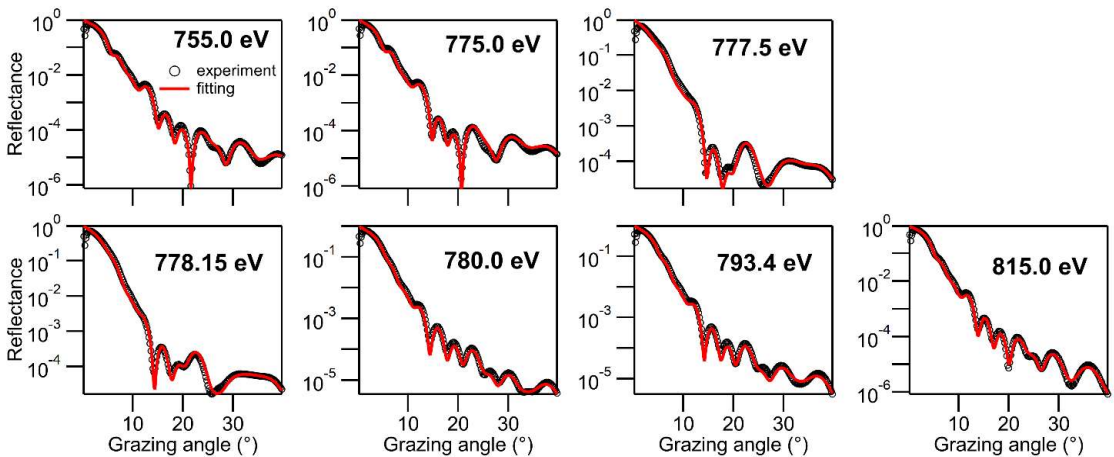

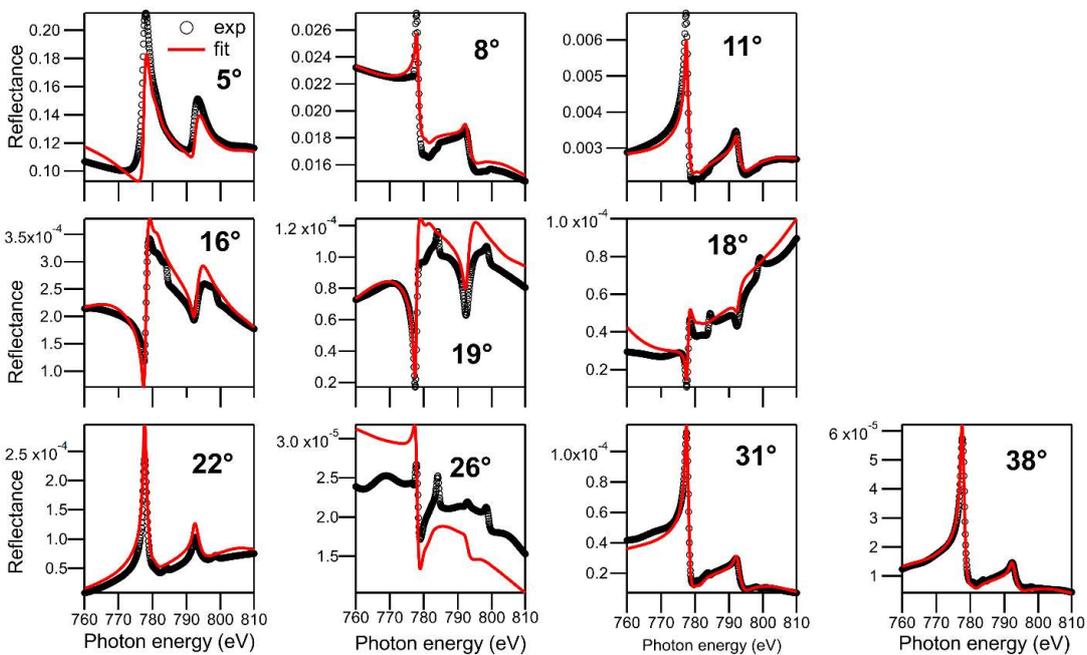

# Figure S10

Dichroic reflectivity spectra acquired on the multilayer sample (empty black circles) and the simulated curves (blue lines) obtained with a uniformly magnetized Co layer, i.e., a Co magnetization profile equal to the Co concentration profile multiplied by a constant. In the top panels: the asymmetry ratio $AR$ acquired in $\theta - 2\theta$ scans at fixed photon energy (775.0-793.4 eV). In bottom panels: the dichroic reflectivity $\Delta R$ acquired in the photon-energy scans at fixed grazing angle (5°-36°).

**θ-2θ scans** (fixed photon energy)

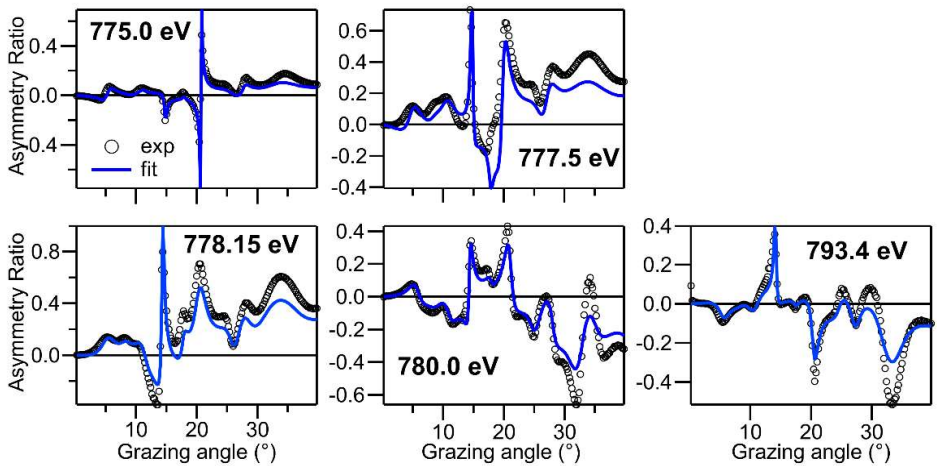

**Photon-energy scans** (fixed grazing angle)

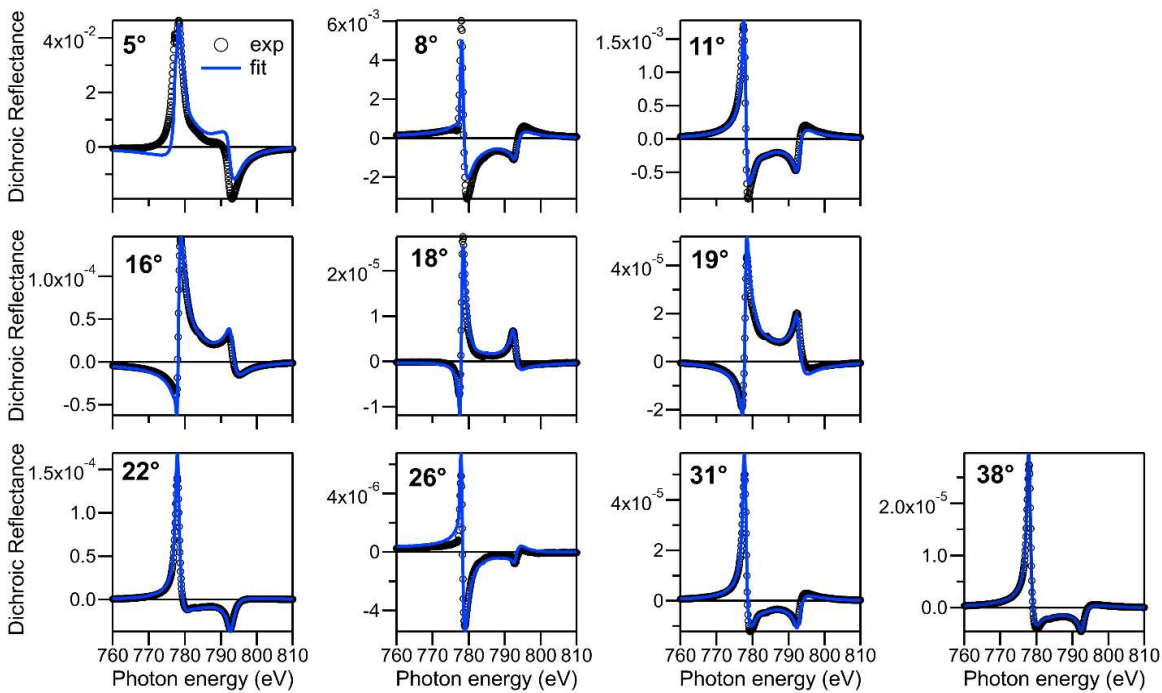

# Figure S11

Dichroic reflectance $\Delta R$ photon-energy spectra across the Pt M$_3$ edge measured on the multilayer sample (black empty circles) at nine values of grazing angle $\theta$ (1.5°-7.0°) and the corresponding curves obtained by the fitting procedures (blue lines) that provides the Pt magnetization depth profile in **Figure 5e**.

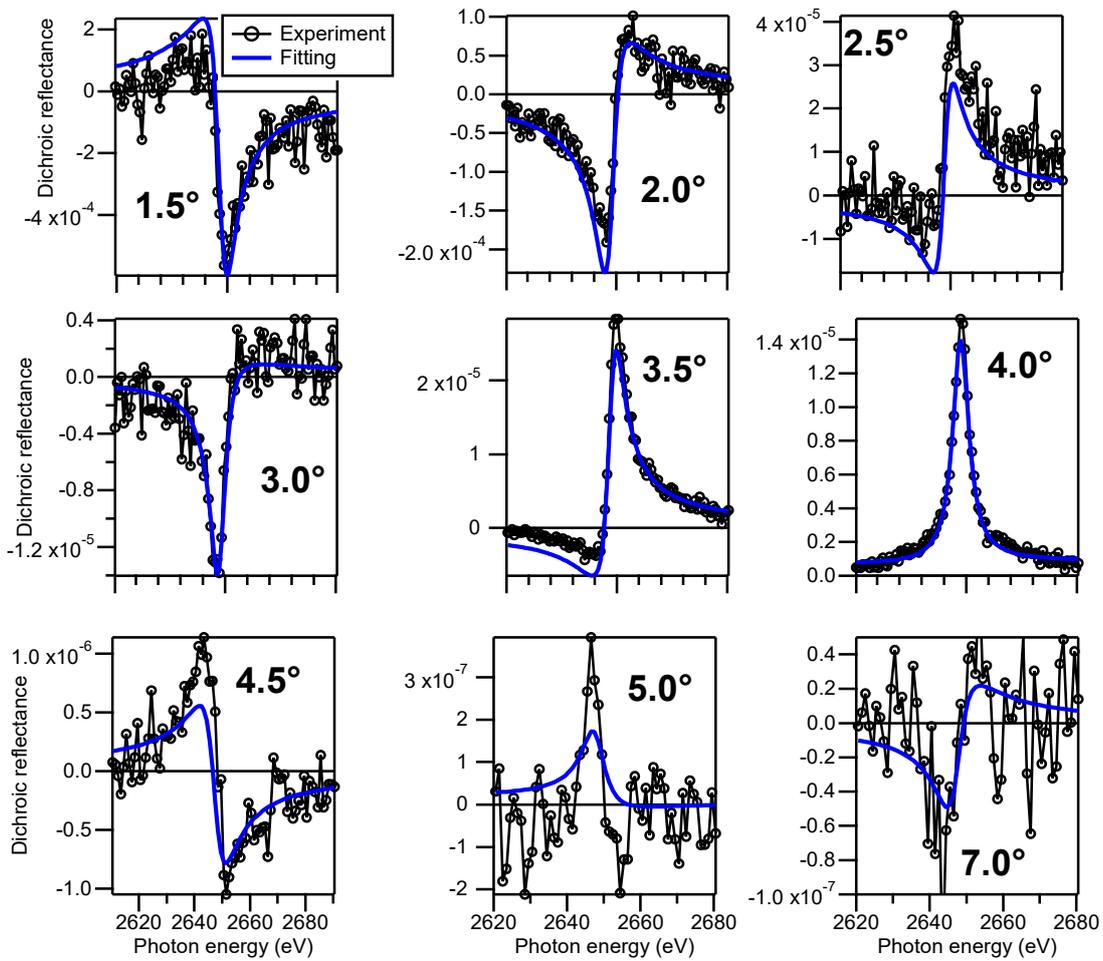

# Figure S12

Resonant X-ray photon-energy spectra across the Co $L_{2,3}$ edge revealed the presence of two unexpected very small peaks in addition to the main Co $L_{2,3}$ structure for both the reference trilayer sample and for the multilayer sample as shown in **Figures S5** and **S9** of the Supporting Information. These additional peaks, not visible in the more surface sensitive TEY, were attributed to the Ba $M_{4,5}$ thresholds.

In order to confirm the hypothesis of barium contamination, X-ray photoelectron spectroscopy (XPS) measurements combined with ion sputtering were carried out on the additional trilayer sample with a 15 Å Pt cap layer, described in the presentation of **Figures 3c-d**. The investigated photoemission lines are nominally: Si 2s (binding energy $E_B$=149.7 eV), O 1s ($E_B$=543.1 eV), Ta 4f ($E_B(4f_{5/2})$=23.5 eV and $E_B(4f_{7/2})$=22.6 eV), ), Pt 4f ($E_B(4f_{5/2})$=74.5 eV and $E_B(4f_{7/2})$=71.2 eV), Co 2p ($E_B(2p_{1/2})$=793.2 eV and $E_B(2p_{3/2})$=778.1 eV), Ba 3d ($E_B(3d_{3/2})$=795.7 eV and $E_B(3d_{5/2})$=780.5 eV), C 1s ($E_B$=284.2 eV).

The area of the photoemission peaks for the various elements is obtained from a fitting procedure. This area constitutes the intensity of the investigated photoemission lines. In the graph we report the intensity of the investigated XPS lines as a function of the number of sputtering processes.

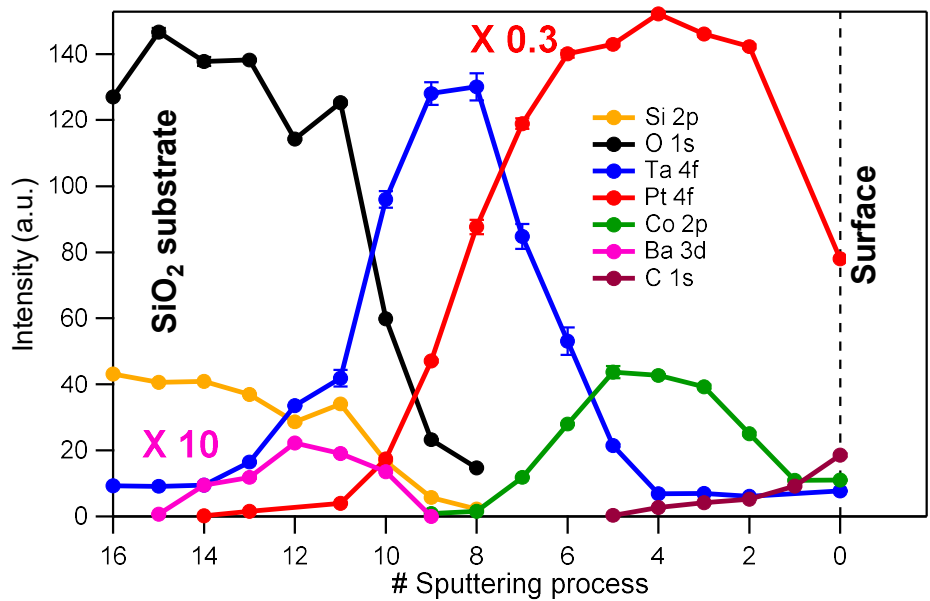

The sputtering process #0 refers to the sample as inserted in the experimental chamber. For a better readability, the curve relative to the Pt 4f peak is multiplied by 0.3 and the curve relative to Ba 3d is multiplied by 10.

The curves reported in the graph provide a qualitative indication of the distribution in depth of the various elements in the sample.

The contamination of Ba atoms appears to be present at the interface between the Ta buffer layer and the $SiO_2$ substrate, but it is absent in the bulk of the substrate. The distribution of barium in the deepest region of the grown film confirms the results expected by the reflectivity measurements, which indicate that Ba must be found at a greater depth with respect to Co and cannot affect the proximity effect at the Pt/Co and Co/Pt interfaces and the magnetic properties of the film. Note that a partial diffusion of Ta within the Pt layer is also observed.

A possible explanation for the presence of barium in the grown heterostructures is the common use of barium oxide in the coating of hot metallic filaments that may be employed in the Si-passivation process. BaO and other metallic oxides are typically used to reduce the work function of the filaments favouring the extraction of electrons by thermionic effect.